\documentclass[a4paper,11pt]{article}
\usepackage{pos}
\usepackage{graphicx,amsmath}
\usepackage{subfigure}
\usepackage{enumitem}
\setlist[itemize]{noitemsep}

\begin{document}
\title{MATHUSLA: A Proposed Long-Lived Particle Detector for High-Luminosity LHC}
\ShortTitle{MATHUSLA}

\author*[a]{Miriam Diamond}
\author[a]{David Richard Curtin}
\author[b]{Erez Etzion}
\author[c]{Caleb Miller}
\author[a]{Runze Ren}
\author[d]{Steven Robertson}
\author[c]{Heather Russell}

\affiliation[a]{Department of Physics, University of Toronto, 60 St George Street, Toronto, Canada}
\affiliation[b]{School of Physics \& Astronomy Department of Particle Physics, Tel Aviv University, 69978 Tel Aviv, Israel}
\affiliation[c]{Victoria Subatomic Physics \& Accelerator Research Centre, University of Victoria, 3800 Finnerty Road, Victoria, Canada}
\affiliation[d]{Department of Physics, University of Alberta,  11335 Saskatchewan Dr NW, Edmonton, Canada}
\emailAdd{m.diamond@utoronto.ca}
\emailAdd{d.curtin@utoronto.ca}
\emailAdd{ereze@tauex.tau.ac.il}
\emailAdd{calebmiller@uvic.ca}
\emailAdd{tom.ren@utoronto.ca}
\emailAdd{steven.robertson@ualberta.ca}
\emailAdd{heather.russell@cern.ch}

\abstract{Beyond the Standard Model long-lived particles (LLPs) appear in many theories that address fundamental open questions, such as the hierarchy problem and the nature of dark matter. The Large Hadron Collider (LHC) may be producing copious numbers of neutral LLPs with masses above a GeV, only for these sneaky particles to escape the main detectors without being spotted. To fill this gap, we have proposed MATHUSLA (MAssive Timing Hodoscope for Ultra-Stable neutraL pArticles), which would be constructed on the surface above CMS and would take data during High-Luminosity LHC operations. The detector would be composed of several layers of solid plastic scintillator threaded with wavelength-shifting fibers that connect to silicon photomultipliers (SiPMs), and would monitor an empty, air-filled decay volume. The Conceptual Design Report (CDR) published in 2025 sets out a benchmark geometry of $40\mbox{m}\times40\mbox{m}\times16\mbox{m}$, with a modular detector construction scheme that would allow data collection to begin as soon as the first module is installed. This proceeding documents the results of more detailed studies conducted since the CDR on the detector's potential location, mechanical engineering of modules, data acquisition system, cosmic-ray backgrounds, event selection criteria, and trigger algorithms.}

\FullConference{43rd International Conference on High Energy Physics - ICHEP2026\\
		Jul 30 - Aug 5, 2026\\
		Natal, Brazil}
        
\maketitle

Beyond the Standard Model, neutral long-lived particles (LLPs) may be produced in copious numbers at the Large Hadron Collider (LHC) interaction points, but would be difficult to detect with ATLAS, CMS, or LHCb. A large fraction would escape the detector before decaying if their mean proper lifetime significantly exceeds the detector size $(~10$m), while backgrounds could bury the signal of LLPs that do decay inside the detector.
Multiple specialized detectors, complementary to each other and to the main detectors, have been approved or proposed to achieve wide coverage of LLP parameter space.
The forward category (e.g. FASER, SHiP, NA62) target LLP production at low centre-of-mass energies, with a relatively small detector volume. The transverse category targets LLP production at high center-of-mass energies, with a much higher detector volume; proposals include MATHUSLA (MAssive Timing Hodoscope for Ultra-Stable neutraL pArticles) as well as ANUBIS \cite{Bauer:2019vqk},  CODEX-b \cite{Aielli:2019ivi}, and GRENDEL \cite{citron2026_grendel}.
MATHUSLA is a surface detector whose long distance from the interaction point provides shielding against SM collision backgrounds, while its large volume ensures sufficient LLP acceptance.
Targeting zero-background searches for LLPs with the lifetimes all the way up to the big bang nucleosynthesis limit ($\sim10^7-10^8\mbox{m}$), MATHUSLA's projected sensitivity in large swaths of LLP parameter space exceeds that of the LHC main detectors by orders of magnitude. Here, we highlight the progress in the detector design and background modeling that has been made since the 2025 Conceptual Design Report \cite{MATHUSLA:2025zyt}.  

\vspace{-2mm}
\paragraph{Conceptual Design Report Summary:}

MATHUSLA's primary physics motivation is the discovery of exotic Higgs decays to long-lived scalars ($H \rightarrow XX$, e.g. the Showcase Model in the Physics Beyond Colliders Summary Report~\cite{PBC}).
Secondary physics motivations include high-mass singlet scalars (SM+S, e.g. PBC Benchmark Case 4), heavy neutral leptons/right-handed neutrinos (RHN, e.g. PBC Benchmark Cases 6,7, and 8), and other exotic modes (e.g. dark glueballs). MATHUSLA is envisioned as a $40\mbox{m}\times40\mbox{m}\times16\mbox{m}$ surface detector above CMS, constructed as sixteen $9\mbox{m}\times9\mbox{m}$ towers. Its modular design allows staged installation, in which more towers are added while existing towers collect data. Each tower features six tracking layers of solid scintillator for reconstructing displaced vertices from LLP decays, two similar floor layers for vetoing SM collision backgrounds, and a decay volume filled with air. Background rates can be independently measured during beam-off periods. MATHUSLA can run completely standalone, though combination with CMS to facilitate the characterization of LLP signals is foreseen.


The tracking layers are made from plastic scintillator bars and threaded by wavelength- shifting optical fibres. Each layer is comprised of ``bar assemblies" that can be manufactured in a modest-sized lab. Each fiber loops through two bars and is coupled to SiPMs at both its ends, such that a hit position along a bar can be derived from the time difference between SiPM signals. The bar assemblies are optimized for low-cost and high-volume production while meeting the requirement of ``4D'' $(x, y, z, t)$ hit information with $\approx 1$~ns timing resolution and $\approx 14$~cm transverse position resolution. This resolution is crucial for distinguishing upward-going from downward-going tracks, which helps with the identification and rejection of cosmic backgrounds. MATHUSLA would save all raw hit data for analyses independent of CMS; the resulting data volume would be $\sim 0.6$~TB/day, $O$(1 PB) for 5 years runtime.
For analyses combined with CMS, MATHUSLA would send a signal to CMS to trigger on potential LLP-producing bunch crossings. For this, a separate ``CMS-LLP'' trigger would have to be developed, targeting a rate of $<1$ Hz. It would have to be issued with latency $<2 ~\mu\mbox{s}$ to arrive at CMS within their L1 latency budget. Since the LLP velocity is unknown, there are many bunch crossings that could be responsible for a given displaced vertex in MATHUSLA, and work with CMS will be required to develop a system to read out the multiple interesting bunch crossings in response to the LLP trigger.

\textsc{MadGraph} and \textsc{PYTHIA} were employed to simulate $H \rightarrow XX, X \rightarrow bb$, SM+S, and RHN signal events. For the backgrounds, LHC muons and neutrinos were generated with \textsc{MadGraph} + \textsc{GENIE} + \textsc{PYTHIA}, atmospheric neutrinos by \textsc{GENIE} + \textsc{PYTHIA}, and cosmic rays by \textsc{PARMA} + \textsc{CRY}. With a full \textsc{GEANT4} detector model, hits from the simulated events were corrected for detector resolution and efficiency, and put through track and vertex reconstructions. 
For $H \rightarrow XX, X \rightarrow bb$ decays, a set of event selection criteria was found that would permit a near zero-background search.

Mini-towers of footprint $\sim 1$~m$^2$ with 4 tracking layers, one at the University of Victoria and one at the University of Toronto, are operating for MATHUSLA hardware optimization, DAQ development and testing, and cosmic ray analyses. Reconstruction of cosmic ray tracks has been demonstrated with high efficiency in both mini-towers.

\vspace{-2mm}
\paragraph{Proposed Detector Location and Civil Engineering:}

\begin{figure}[h]
\centering
\includegraphics[width=\textwidth, trim=0 40mm 0 0, clip]{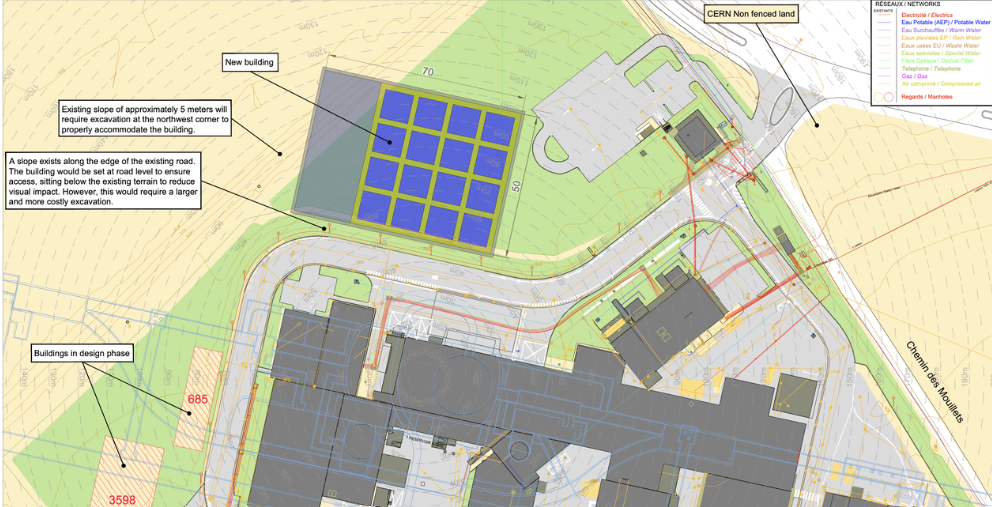}
\caption{Concept drawing for MATHUSLA detector hall, containing 16 towers (blue squares), near CMS. Yellow land is ``CERN non-fenced'', green is CERN fenced. (Credit:  CERN Site \& Civil Engineering Dept.) }
\label{fig:building-location}
\end{figure}

\begin{figure}[h]
\centering
\includegraphics[width=\textwidth]{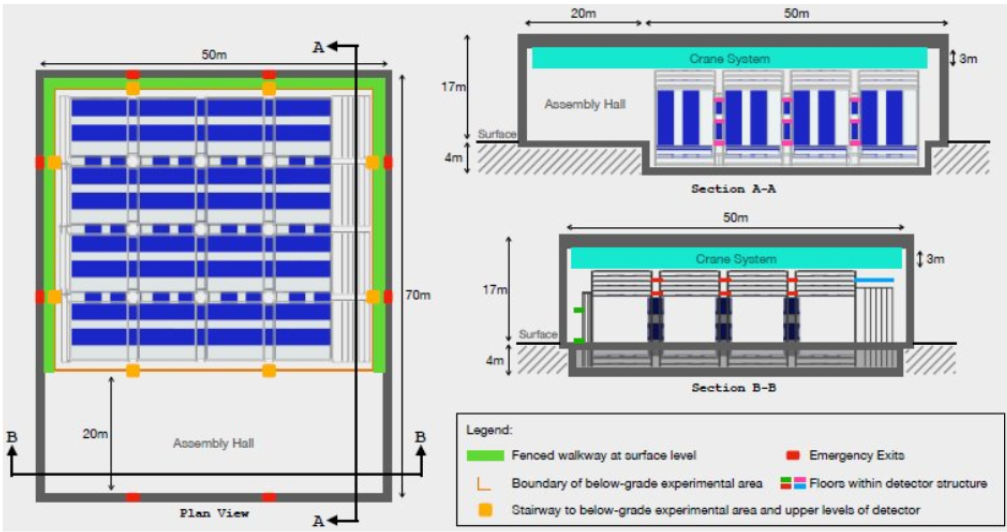}
\caption{Concept drawing for MATHUSLA detector hall. (Credit:  CERN Site \& Civil Engineering Dept.)}
\label{fig:building-engineering}
\end{figure}

Figure~\ref{fig:building-location} shows a possible location of the MATHUSLA ``detector hall'' near CMS. CERN determined that sites near ATLAS would be logistically problematic, due to access and congestion issues. Note that the shape of the hall is not fixed, as the towers do not have to be arranged in a $4\times4$ square, nor even in a rectangle. The relative positions of the towers does not significantly impact overall the physics reach. A possible layout for the detector hall itself, in the simplest case of tower arrangement (square), is shown in Figure~\ref{fig:building-engineering}.

\vspace{-2mm}
\paragraph{Engineering of Bar Assemblies:}

\begin{figure}[h]
\centering
\includegraphics[width=0.6\textwidth]{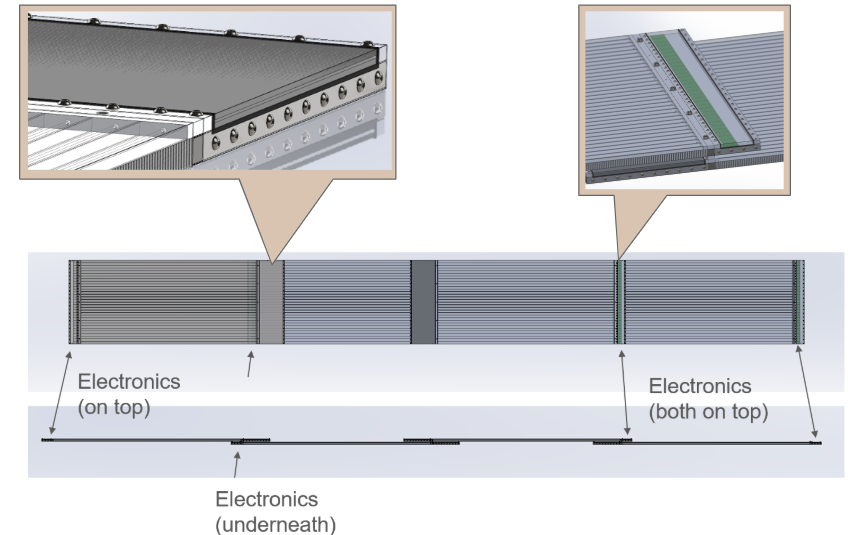}
\includegraphics[width=0.48\textwidth]{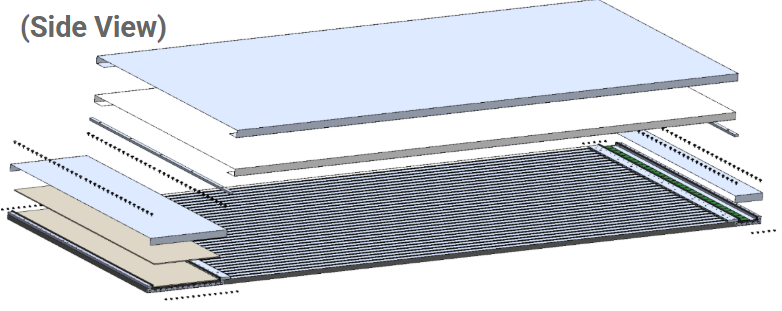}
\includegraphics[width=0.48\textwidth]{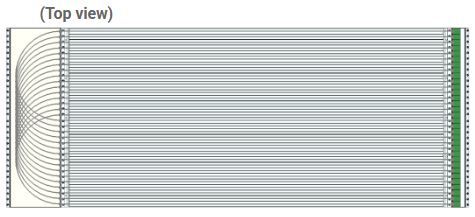}
\caption{Bar assemblies, engineering drawings. (Credit: Mitchel Baker, NSERC Major Resources Support).}
\label{fig:bar-assemblies}
\end{figure}

Much progress has been made in the detailed bar assembly design (see Figure~\ref{fig:bar-assemblies} for engineering drawings). Open questions include: the sufficiency of industrial-grade, double-sided tape to mount the scintillator bars to aluminum honeycomb panel on one side and sheet aluminum on the other; the minimum arrangements of threaded inserts and screws required; making the instrumentation and fiber-loop areas accessible for servicing, but light-tight when closed; and shielding exposed instrumentation cabling from damage.

\vspace{-2mm}
\paragraph{Data Acquisition System:}

\begin{figure}[h]
\centering
\includegraphics[width=0.45\textwidth]{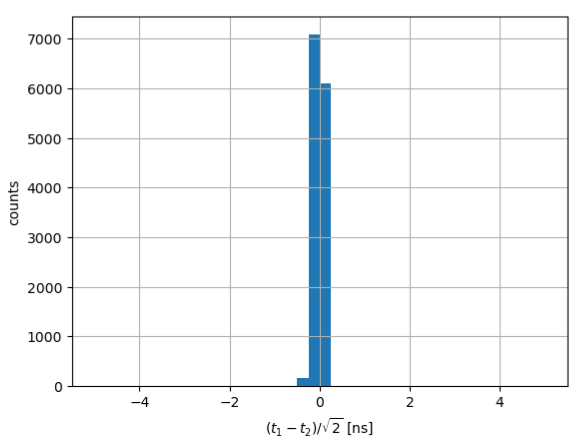}
\includegraphics[width=0.48\textwidth]{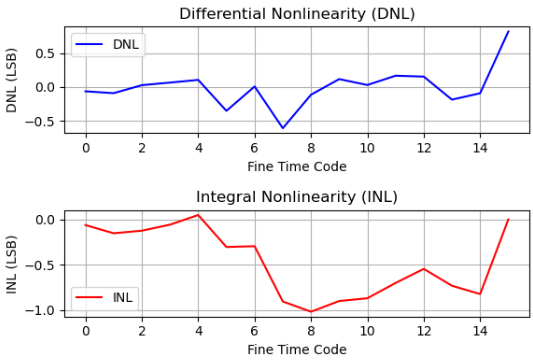}
\caption{Timing resolution of 0.09\,ns and 12\,ns (3 clock cycles) latency demonstrated with multi-phase clock sampling TDC implemented directly in FPGA.}
\label{fig:DAQ-plots}
\vspace{-2mm}
\end{figure}

The University of Toronto mini-tower uses a commercial ASIC (TOFPET2) with a custom front-end board. The SiPM preamplifier is custom-made with Monolithic Microwave Integrated Circuit components. The ASIC has 64 channels of integrated time readout; the resolution and throughput is sufficient, but the  latency is too high for implementing the CMS-LLP trigger. The latency is due to the time to digital converter (TDC), which can achieve ps-scale resolution but suffers from delays from the conversion mechanism and the overhead of the Serializer/Deserializer blocks. We are therefore switching to multi-phase clock sampling TDC implemented directly in FPGA, with timing resolution results shown in Figure~\ref{fig:DAQ-plots}.

\vspace{-2mm}
\paragraph{Background Simulations and Event Selection:}
As \textsc{PARMA} and \textsc{CRY} use parametric models, an additional cosmic background study was performed using \textsc{CORSIKA} to investigate the correlated effects of multiplicity in cosmic showers.
At higher energies, events with multiple protons and/or neutrons are typically correlated with large showers of charged particles, which allows for an effective veto of these multi-hadron events. Lower-energy primaries that produce multi-hadron events have low spatial correlation, so should be sufficiently modeled by the multi-baryon injection in \textsc{PARMA} and \textsc{CRY}.

To optimize the event selection criteria, analysis was re-run on all the simulated signal and background events with two different requirements on the minimum number of tracks required to construct a displaced vertex. For three tracks (four tracks), LLP signal efficiency was 40-50\% (30-40\%), with a mean of 1.34 (0.23) background events per year of full detector runtime. Table~\ref{tab:event-selection} shows the 90\% CL upper limits on events from each background source category. 
 \begin{table}[]
     \centering
    \begin{tabular}{|c|c|c|c|}
    cosmic p, n & LHC $\mu$ & Atmospheric $\mu$ & LHC $\nu$ \\
          \hline
    3.04 & 0.36 & 0.11 & 0.06  \\
    \end{tabular}
     \caption{Expected background events (90\% CL upper limit) in five years of full detector runtime, after applying event selection criteria, with a minimum of three tracks required to construct a displaced vertex.}
     \label{tab:event-selection}
 \end{table}

\vspace{-2mm}
\paragraph{Trigger Development:}
The goals for the CMS-LLP trigger are to keep as many MATHUSLA-reconstructable physics events as possible while reducing the selection rate of background cosmic events to $<1$~Hz, and to
use primitive information as much as possible to achieve the $<2 \mu$~s latency on a minimum number of parallel FPGAs. Fast approaches to trigger algorithms employ hit counts and tower/layer activity, while slow approaches employ multi-layer track and vertex reconstruction; in-between are 2-hit pairs and the direction of the evolution of hits within an event. The trigger we developed uses fast approaches in events with high occupancy and slower approaches only for lower-occupancy, ambiguous events. Our algorithm achieves high acceptance of reconstructible events (Table~\ref{tab:trigger}) with a cosmic rate of 0.8~Hz; several parameters can be tuned to further reduce this rate if necessary. 
We now investigating the algorithm's implementation in FPGA: the latency, the required specifications for the FPGA itself, and whether events with $\geq 200$ hits can be handled in the same pipeline as events with $O(1)$ hit. 

\begin{figure}[h]
\centering
\includegraphics[width=\textwidth]{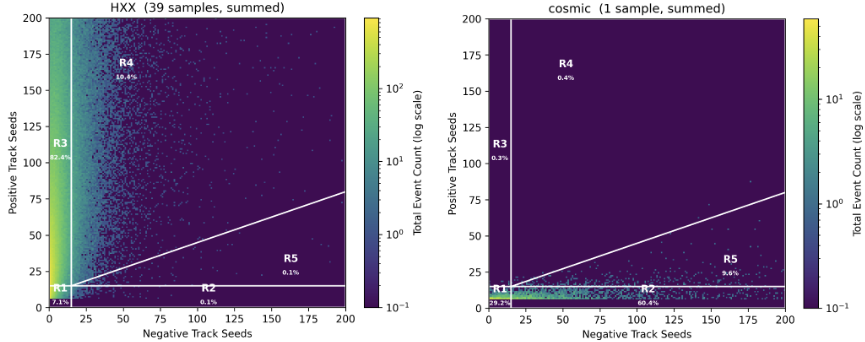}
\caption{Counts of positive (upward-going) vs negative (downward) track seeds in simulation (one sample = 10000 events), divided into regions R1 through R5. Left: $H\rightarrow XX$,$X \rightarrow bb$ signal events. Right: cosmics.}
\label{fig:trigger}
\end{figure}

To develop the algorithm, the numbers of upward- and downward-going ``track seeds'' (pairs of spatially-close, causally-linked hits) were found in the simulated events. Events were then divided into ``regions'' (Figure~\ref{fig:trigger}) based on the 2D distribution of upward- vs downward-going seeds. The algorithm goes as follows: events in Regions 3 or 4 are kept, while in the more ambiguous Region 1, tracks with $\geq 4$ hits are reconstructed starting with the track seeds; track compatibility with a common vertex is checked if at least two valid tracks are found; and any event with a vertex  within the decay volume is triggered.
Accepting only events falling into Regions 3 or 4 keeps $>= 90$\% of $H \rightarrow XX, X \rightarrow bb$ signal events with a sufficiently low cosmic background rate.

 \begin{table}[]
     \centering
    \begin{tabular}{c|c|c}
          Signal & For All Events in Detector & For Reconstructible Events \\
          \hline
          $H \rightarrow XX$, $X \rightarrow bb$& 93-100\% & $\sim 100$\%  \\
          Right-handed neutrino & 32-63\% & 86-97\% \\
          Singlet heavy scalar & 3-47\%& 44-91\% \\
          \hline
    \end{tabular}
     \caption{Efficiency of the trigger algorithm applied to simulated ($\sim 1$\% stat. uncertainty) LLP signal events.}
     \label{tab:trigger}
 \end{table}

\vspace{-7mm}

\bibliographystyle{JHEP}
\bibliography{references2}

@article{MATHUSLA:2025zyt,
    author = "Aitken, Branden and others",
    title = "{Conceptual Design Report for the MATHUSLA Long-Lived Particle Detector near CMS}",
    eprint = "2503.20893",
    archivePrefix = "arXiv",
    primaryClass = "physics.ins-det",
    reportNumber = "FERMILAB-PUB-25-0314-PPD",
    month = "3",
    year = "2025"
}

@article{citron2026_grendel,
      title="{Proposing GRENDEL: GalleRy ExperimeNt for Decays of Exotic LLPs at the LHC}", 
      author={Matthew Citron and Federico Redi and Ryan Schmitz},
      year={2026},
      eprint={2609.00152},
      archivePrefix={arXiv},
      primaryClass={hep-ex},
      url={https://arxiv.org/abs/2609.00152}, 
}

@article{Bauer:2019vqk,
    author = {Martin Bauer and Oleg Brandt and Lawrence Lee and Christian Ohm},
    title = "{ANUBIS: Proposal to search for long-lived neutral particles in CERN service shafts}",
    eprint = "1909.13022",
    archivePrefix = "arXiv",
    primaryClass = "physics.ins-det",
    month = "9",
    year = "2019"
}

@techreport{PBC,
    author={R. Alemany Fernández and others},
    title="{Summary Report of the Physics Beyond Colliders Study at CERN}",   
    institution="{CERN}",
    number = "CERN-PBC 2025-003",
    eprint={https://cds.cern.ch/record/2927631},
    year={2025},
    month={June}
}

@article{Aielli:2019ivi,
    author = "Aielli, G",
    title = "{Expression of interest for the CODEX-b detector}",
    eprint = "1911.00481",
    archivePrefix = "arXiv",
    primaryClass = "hep-ex",
    doi = "10.1140/epjc/s10052-020-08711-3",
    journal = "Eur. Phys. J. C",
    volume = "80",
    number = "12",
    pages = "1177",
    year = "2020"
}

\end{document}